\documentclass[a4paper,12pt]{article}
\usepackage{amsmath,amssymb}
\usepackage[cp1251]{inputenc}
\usepackage{graphicx,color}

\begin{document}
\title{Phase transition in liquids with directed
intermolecular bonding}%
\author{R.Ryltcev, L.Son}
\date{}
\maketitle
Ural State Pedagogical University, 620219 Ekaterinburg,
Russia.\\


\begin{abstract}
Liquids with quasi - chemical bonding between molecules are
described in terms of vertex model. It is shown that this bonding
results in liquid - liquid phase transition, which takes place
between phases with different mean density of intermolecular
bonds. The transition may be suggested to be a universal phenomena
for those liquids.
\end{abstract}

{\bf keywords:}~Liquid - liquid phase transition, intermolecular
bonds, vertex model.

\section{Introduction.}

The idea that  bonding in molecular liquids may result in
existence of two structural states, was first suggested for water
\cite{stan1}-\cite{stan4}. At low temperatures, supercooled water
demonstrates two possible local arrangements of molecules at some
intermediate length scale, high density (HD) and low density (LD)
ones. This results in a competition between these two states, so
the supercooled water may be treated as a binary mixture of HD and
LD liquids \cite{Ponyatovskii}. The liquid - liquid phase
transition \cite{stan1,Ponyatovskii} in supercooled water is the
ultimate manifestation of the HD and LD local states. Now, the
water HD and LD local structures are well established both in
experiment and in computer simulations (see,for example,
\cite{Malenkov}). Intermolecular coupling by hydrogen bonds occurs
to be a mechanism for HD and LD local states formation
\cite{stan3}, original idea was published in \cite{stan4}.

Recently, the unexpected physical behavior was found in Benzene,
Quinoline, and other complex fluids \cite{lq1}-\cite{lq8}. Namely,
the anomalies (discontinuities, kinks, peaks) on temperature
dependencies of a variety of physical properties were observed.
These anomalies may be treated as a manifestation of changes in
the local structure and may be observed, under special
experimental conditions, in many equilibrium and supercooled
liquids. The molecules of those liquids demonstrate some common
features. For considerations related to the structure of the
liquid, these molecules may be treated as rigid bodies without
internal degrees of freedom. The relatively rigid electronic
configurations of the molecules are characterized by a significant
separation of positive and negative electric charges at the
molecular length scale. At very short distance between molecules,
strong repulsion determines the interaction, and at the only
slightly larger length-scale, attraction become important. At this
distance, the electric charge distribution may be characterized in
terms of rigid electrical multipoles. Interactions of these
multipoles, together with the very short-range repulsion,
determine the set of mutual positions and orientations of two
neighboring molecules corresponding to potential energy minima,
which provide intermolecular bonds.

Thus, the intermolecular coupling seems to be a mechanism for two
or more structures formation at some intermediate length scale. In
present paper, we apply known method of description of the polymer
solution to the statistics of couplings in molecular liquids. The
structure of paper is as follows.
\begin{itemize}
\item In section 2, we give a brief introduction into formalism
applied.
\item Then, in section 3, we analyze the mean - field solution of
the model, which demonstrates liquid - liquid phase transition.
\item In section 4, exact solution of the model
on the Bethe lattice is analyzed. It is shown that the phase
transition takes place in this case too.
\item In Conclusion, general evaluation of the model is given.
\end{itemize}
The aim of the paper is to demonstrate that local structure
separation is rather universal phenomena for liquids with hydrogen
- like bonding.

\section{The model.}

Statistical description of condensed substances with the
possibility of association and aggregation of molecules has been
intensively developed for a long time. The theory of associated
solutions \cite{Prigogine}, the Flory theory of polymer solutions
\cite{Flory},  are the milestones on this way. The idea of
modelling of chemically aggregated systems in terms of scalar
order parameter has a long history also. Probably I.Lifshitz was
first who offered this approach \cite{Lifshitz}. Then, Flory mean
- field theory \cite{Flory} and spin - polymer analogy
\cite{DeGennes} were applied to describe polymer systems. In the
present paper, we consider the model which is a step on this way.

Quasi - chemical  bonding between molecules has been characterized
by two main features. First, the bond is short - range and
directed (i.e., may be drawn as the vector connecting two
neighboring molecules). Second, the bonding is saturated, i.e. no
more than some characteristic number of bonds $N$ may be permitted
per one molecule. The same features may be suggested for the bonds
in polymers, so the statistical description of polymer solutions
may be applied to the molecular liquids with bonding. The
difference is that in the theory of polymers, one supposes the
polymer chain very long, while in our case, the chain of bonded
molecules may have an arbitrary length.

The idea of the model arises from the works of Baxter
\cite{Baxter}, Nikomarov and Obukhov \cite{Obukhov}, and may be
formulated as follows.

Consider spatial net of sites which approximately corresponds to
the spacing of molecules in the liquid. Neighboring sites are
connected with each other by edges. The number of nearest
neighbors we denote as $\gamma$. Let us numerate the sites by
Greece indexes, and the edges by Latin ones.
 Each site $\alpha$ is characterized by  variable $n_\alpha$,
 which is the filling number
\begin{equation}\label{na}
n_\alpha=\left\{
\begin{array}{ll}
 1, & \textrm{if the site $\alpha$ is occupied by the molecule}\\
 0, & \textrm{if the site $\alpha$ is empty}\\
\end{array} \right.
\end{equation}
Also, real scalar variable $\psi_i$ corresponds to the edge
numbered by $i$. Let us consider the partition
\begin{equation}\label{Z}
Z=\sum_{\{n\}} \int D\psi \exp[-F\{n,\psi\}],
\end{equation}
where the summation and integration goes over all possible
configurations $\{n,\psi\}$ of variables $n_\alpha, \psi_i$. The
"effective Hamiltonian" $F\{n,\psi\}$ is defined by
\begin{eqnarray}\label{F}
F\{n,\psi\} =  - \frac{1}{2}\sum_{\alpha,\beta}n_\alpha J_{\alpha
\beta}n_\beta-\frac{\mu}{T} \sum_\alpha n_\alpha +
\frac{1}{2}\sum_i K \psi _i^2- \sum_\alpha \ln \left[ 1 + n_\alpha
R_\alpha(\psi)  \right] \\
R_\alpha(\psi)=a_1\sum_{i^1_\alpha}\psi_{i^1_\alpha}+a_2\sum_{i^1_\alpha
\neq
i^2_\alpha}\psi_{i^1_\alpha}\psi_{i^2_\alpha}+\ldots+a_N\sum_{i^p_\alpha
\neq
i^q_\alpha}\psi_{i^1_\alpha}\psi_{i^2_\alpha}\ldots\psi_{i^N_\alpha}.
\label{R}
\end{eqnarray}
Here, indexes $i^1_\alpha, i^2_\alpha,\ldots i^p_\alpha,$ numerate
the edges which are connected with the site $\alpha$. Spatial
matrix $J_{\alpha \beta}$ is
\begin{equation}\label{J}
J_{\alpha \beta}=\left\{
\begin{array}{ll}
 J/T, & \textrm{if $\alpha , \beta$ are the nearest neighbors}\\
 0, & \textrm{otherwise}\\
\end{array} \right. ,
\end{equation}
where $J$ is the energy of attractive non - directed (Van der
Waals) interaction between molecules, and $\mu$ is their chemical
potential. Thus, the first and the second terms in (\ref{F})
correspond to the lattice - gas model \cite{Baxter} and allow one
to describe the density of the system. The value $K$ is
\begin{equation}\label{K}
K=\exp(U/T),
\end{equation}
where $U$ is the energy of quasi - chemical bond. In (\ref{J},
\ref{K}), $T=k_B T_k$, where $T_k, k_B$ are the temperature and
the Boltzmann constant respectively. The last two terms
in(\ref{F}) allow one to describe bonding. To demonstrate this
fact, let us rewrite (\ref{Z})
\begin{eqnarray}
Z&=&\sum_{\{n\}}e^{-F\{n\}} \int D\psi
\exp\left[-\frac{1}{2}\sum_i K \psi
_i^2\right]\prod_\alpha\left(1+n_\alpha R_\alpha(\psi)\right), \nonumber\\
F\{n\}&=&- \frac{1}{2}\sum_{\alpha \beta}n _\alpha
J_{\alpha\beta}n_\beta-\frac{\mu}{T} \sum_\alpha n_\alpha,
\end{eqnarray}
and consider the functional integral over $\{\psi\}$ at some given
configuration of filling numbers $\{n\}$. Releasing the brackets
in the product, one gets the sum over all possible series of
products:
\begin{equation}\label{Z1}
\prod_\alpha\left(1+n_\alpha R_\alpha(\psi)\right)= 1+n_1 a_1
\psi_{1_1} + n_1 a_1 \psi_{2_1}+ \ldots n_1 n_2 a_1^2
\psi_{1_1}\psi_{2_1} +\ldots .
\end{equation}
Then, the integration with the Gaussian weight
$\exp\left[-\frac{1}{2}K\sum \psi_i^2\right]$, produces all
possible couplings between pairs of neighboring sites. The
coupling takes place when the mathematical power of $\psi$ on
corresponding edge is two. Zero power corresponds to the edge
without coupling. The ratio of the weight of coupled edge to the
weight of uncoupled one is $K^{-1}=\exp(-U/T)$. Thus, the
partition (\ref{Z1}) generates the sum over all possible
configurations of occupied sites, connected by all possible
couplings between nearest neighbors, which correspond to
intermolecular bonds. Each site with $k$ bonds along its edges has
an additional weight $a_k$. No more than one bond per edge, and no
more than $N$ bonds per site are permitted due to the structure of
polynomial $R_\alpha(\psi)$, see (\ref{R}).

 Besides, each configuration of filling numbers
$\{n\}$ has the standard weight of the lattice - gas theory,
$\exp[\frac{1}{2}\sum_{\alpha \beta}n _\alpha
J_{\alpha\beta}n_\beta-\frac{\mu}{T} \sum_\alpha n_\alpha]$. Thus,
in the frames of formalism considered, one can easily model an
arbitrary molecular liquid by an appropriate choice of model
parameters. Those are:
\begin{center}
\begin{tabular}{|r|l|}
\hline $U$ & The energy of quasi - chemical bond \\ $N$ & The
maximal number of bonds per molecule \\ $J$  &
The energy of non - directed (Van der Waals) interaction\\
$\mu$ & The
chemical potential of molecules\\
$a_m$ & The weight of $m$ - bonded molecule\\
\hline
\end{tabular}
\end{center}

Instead of filling numbers, it is more useful to deal with scalar
variables without any limitations imposed. To do so, Hubbard -
Stratonovich transform to conjugated field may be used. Then, the
summation over filling numbers leads to
\begin{eqnarray}\label{Z2}
Z&=&\int D\psi D\varphi \exp\left[-F\{\varphi,\psi\}\right]
\nonumber\\
F\{ \varphi,\psi \} &=& \frac{1}{2} K\sum_i \psi_i^2
+\frac{1}{2}\sum_{\alpha \beta}\varphi_\alpha J_{\alpha
\beta}^{-1} \varphi_\beta - \sum_\alpha \ln
\left[1+e^{\varphi_\alpha+\frac{\mu}{T}}\left(1+R_\alpha(\psi)\right)\right].
\end{eqnarray}
To analyze functional integral (\ref{Z2}), one can apply
traditional method: to find most probable configuration which
minimizes the $F\{\varphi,\psi\}$, then to investigate
fluctuations around it, etc. In the next section, we describe the
first step, which is the mean - field approximation.

\section{Mean - field approximation.}

The mean - field configuration $<\varphi_\alpha>, <\psi_i>$
provides the minima of (\ref{Z2}) and obeys following equations:
\begin{equation}\label{mf1}
\frac{\partial F}{\partial \psi_i}\mid_{<\varphi_\alpha>,
<\psi_i>}=0,~ \frac{\partial F}{\partial
\varphi_\alpha}\mid_{<\varphi_\alpha>, <\psi_i>}=0.
\end{equation}
Instead of $\varphi_\alpha$, it is better to use the variable
$$
q_\alpha=\sum_\beta J_{\alpha\beta}^{-1}\varphi_\beta,
$$
because its mean - field value,
\begin{equation}
w_\alpha=<q_\alpha>,
\end{equation}
coincides with the mean value of filling number $<n_\alpha>$ and
should be understood as the mean concentration of the molecules.

For homogeneous system, $<\psi_i>=\Psi,~w_\alpha=w$, equations
(\ref{mf1}) may be written in explicit form
\begin{eqnarray}\label{e1}
w&=&\frac{\exp[\frac{J\gamma}{T}w+\frac{\mu}{T}+\ln(1+R(\Psi))]}
{1+\exp[\frac{J\gamma}{T}w+\frac{\mu}{T}+\ln(1+R(\Psi))]}\\
\label{e2} \Psi&=&\frac{n_s}{n_e} e^{-\frac{U}{T}} w
\frac{R~'(\Psi)}{1+R(\Psi)}\\
\label{e3} R(\Psi)&=& a_1 \gamma \Psi
+a_2\frac{\gamma(\gamma-1)}{2}\Psi^2+\ldots+a_N C_\gamma^N\Psi^N
\end{eqnarray}
where $n_s/n_e$ is the ratio of the number of sites to the number
of edges in the lattice and $\gamma$ is the number of nearest
neighbors. Solutions of system (\ref{e1}-\ref{e3}) determine the
equilibrium values of $\Psi, w$. For nonequilibrium $\Psi, w$,
expression (\ref{Z2}) gives the density of thermodynamic potential
in the Landau theory \cite{Landau}:
\begin{equation}\label{F3}
f(\Psi,w)=\frac{n_e}{2n_s}e^{\frac{U}{T}}\Psi^2+\frac{J\gamma}{2T}w^2-\ln
\left[1+e^{\frac{J\gamma}{T}w+\frac{\mu}{T}}(1+R(\Psi))\right].
\end{equation}
Solutions of system (\ref{e1}-\ref{e3}) correspond to different
phases of the system. The stable phases correspond to  minimums of
(\ref{F3}).

The first of equations (\ref{e1}) is exactly the equation which
arises in the lattice - gas theory of critical point:
\begin{equation}\label{first}
\ln\frac{w}{1-w}=\frac{J\gamma}{T}w+B,~~B=\frac{\mu}{T}+\ln(1+R(\Psi)).
\end{equation}
Depending on parameters $T,B$, this equation may have from one to
three solutions on the interval $(0,1)$. The solution with
$w\simeq 1$ corresponds to the condensed phase (liquid), while the
solution $w\simeq 0$ represents the gas. We are interested in the
condensed phase behavior, far from the liquid - gas critical
point. Thus, we consider the case when the equation (\ref{first})
has only one solution, $w\simeq 1$. This may always be provided by
an appropriate choice of parameter $\mu$ in (\ref{first}). If the
equation (\ref{first}) has only one solution, then one gets the
Landau theory with single order parameter $\Psi$ and with
thermodynamic potential
\begin{equation}\label{F4}
f(\Psi)=\frac{1}{2b}\Psi^2+\frac{a}{2}w^2(\Psi)-\ln
\left[1+e^{aw(\Psi)+\frac{\mu}{T}}(1+R(\Psi))\right],
\end{equation}
where $w(\Psi)$ is given by (\ref{first}),
$b=\frac{n_s}{n_e}e^{-\frac{U}{T}},~a=\frac{J\gamma}{T}$. Function
$f(\Psi)$ may have several minima, which corresponds to different
local arrangements of bonds. These minima are determined by
equation (\ref{e2}):
\begin{equation}\label{second}
\Psi  = b w(\Psi) \frac{R~'( \Psi)}{1 +R(\Psi)}.
\end{equation}

The formalism considered allows to describe a wide variety of
intermolecular bonding in a liquid. The bonding has been modelled
by the choice of the polynomial under the logarithm in (\ref{F3}):
\begin{equation}
R(\Psi)= a_1 \gamma \Psi
+a_2\frac{\gamma(\gamma-1)}{2}\Psi^2+\ldots+a_N C_\gamma^N\Psi^N
\end{equation}
The number $N$ is the maximal number of bonds per molecule, and
the coefficient $a_m$ is the weight of $m$-bonded molecule. Note,
that $P(\Psi)$ is positive for most physical cases. Since the
coupling arises due to the charge redistribution in the molecule,
the maximal number of bonds is even. The odd number of bonds means
that the charge distribution in the molecule is not symmetric, so
$$
a_{2n}\geq a_{2n-1}.
$$
It can be easily shown, that this provides positive values of
$R(\Psi)$ at any $\Psi$.

At low temperatures, $w(\Psi)\simeq 1$, and (\ref{second}) may be
rewritten as
\begin{equation}
B\Psi  =
\frac{R~'(\Psi)}{1+R(\Psi)},~~B\sim\exp\left[\frac{U}{T}\right].
\end{equation}
Its graphic solution is presented on fig.1a. The right side of the
equation,
$$
\frac{R~'(\Psi)}{1+R(\Psi)}_{\Psi\longrightarrow\pm\infty}\longrightarrow
0,
$$
with some intermediate behavior, as it presented on fig.1a. Since
$U<0$, the left side is a line which slope lowers with temperature
lowering. Thus, if the temperature is low enough, there always
exist nontrivial set of solutions, which starts from the minimum
of $f(\Psi)$ at large negative $\Psi$, and completes with another
minimum at large positive $\Psi$.

Here, some comments should be done. The mean - field approximation
does not answer the question what is the physical difference
between these two low temperature structures. To do so, one needs
higher approximations. Similar situation takes place for the Ising
model in two dimensions \cite{Baxter}. (In 2D, the Ising model may
be formulated in terms of scalar field which corresponds to
magnetization. At the same time, it may be understood as a
statistics of self - closed contours which divide areas with
inverted spins. Mean magnetization gives poor information about
contours). Nevertheless, some assertions may be done. In our case,
absolute value of $\Psi$ is proportional to the mean length of
quasi - polymers \cite{Obukhov}.  The  structure $\Psi<0$ is
metastable, i.e. $f(\Psi<0)> f(\Psi>0)$ (fig.1b). It may be
stabilized by external field $s$ conjugated to $\psi$, i.e. by
adding the term $s \psi$ to the Hamiltonian. Such a term produces
spatial points, where the quasi - polymers have been terminated.
Thus, the difference between $\Psi>0$ and $\Psi<0$ structures lies
in the statistics of polymer tails. For the Bethe lattice, the
metastable state does not exist (see the next section). Since for
the Bethe lattice closed paths are prohibited, then it is natural
to suggest that the metastable state may be characterized by a set
of self - closed quasi - polymers.

\begin{figure}[h]\label{Sol_1}
   \centering
 \includegraphics[width=1.0\textwidth,clip,viewport= 0 0 560 200]{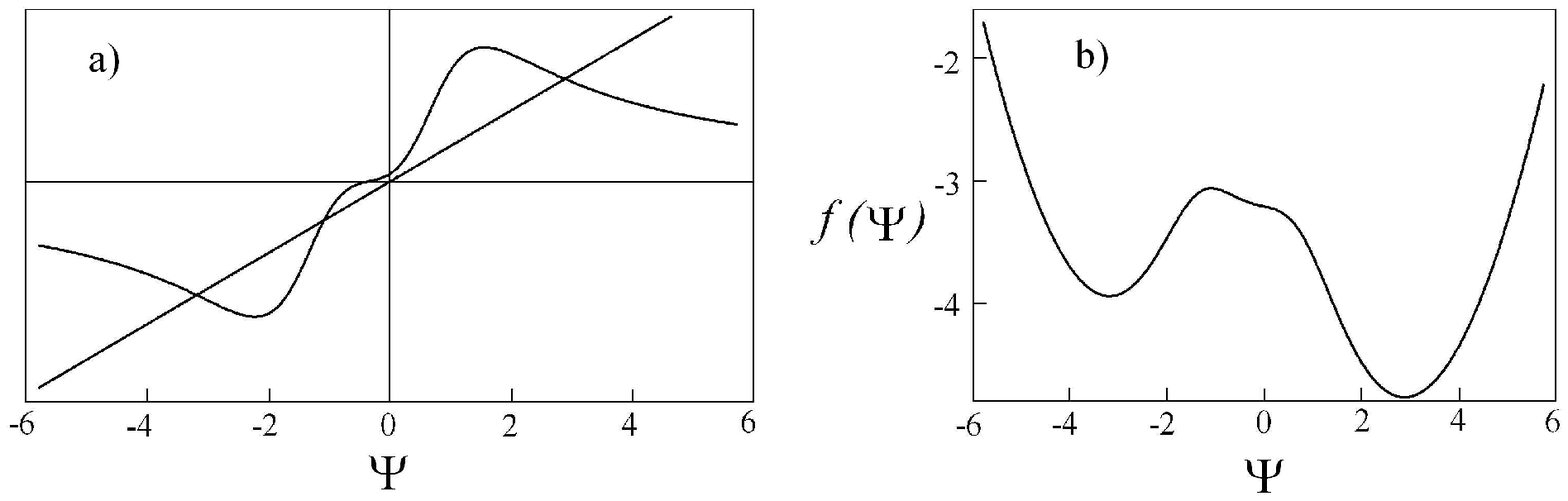}
   \hfil
  \includegraphics[width=1.0\textwidth,clip,viewport= 0 0 545 175]{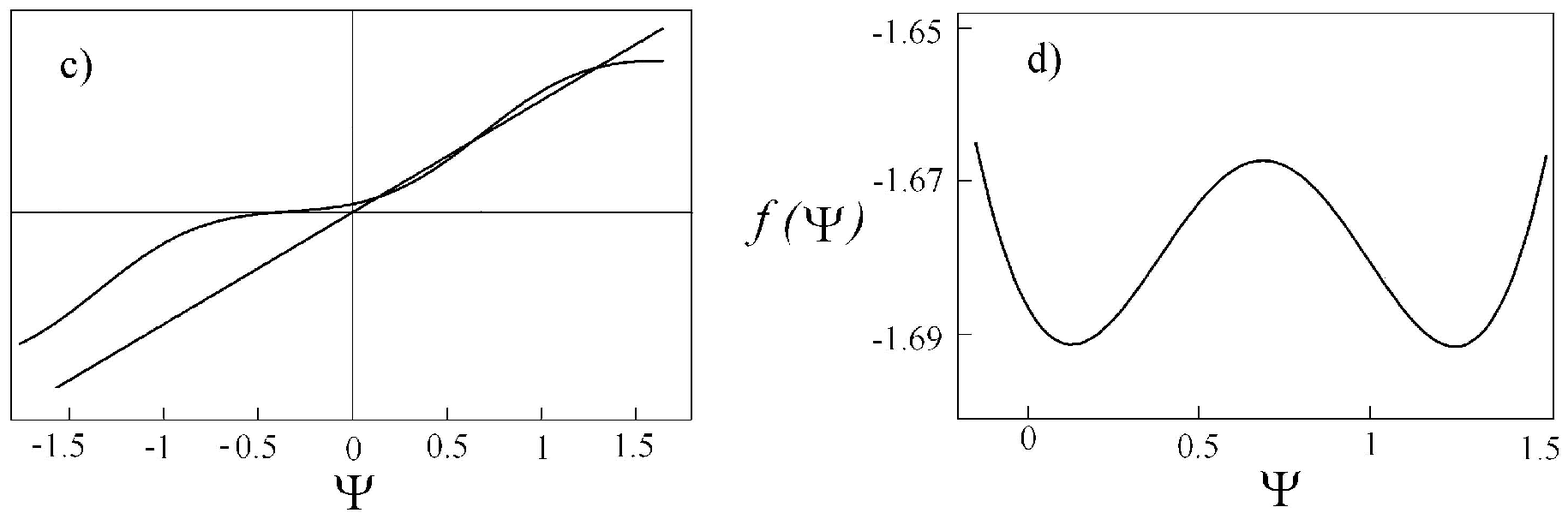}
\hfil   \caption{The graphic solution of equation (\ref{second}),
i.e. left and right sides of this equation, and corresponding plot
of thermodynamic potential (\ref{F4}) at $\tilde T=T/U=0.5$ (a -
graphic solution, b - thermodynamic potential) and $\tilde
T=T/U=1.019$ (c - graphic solution, d - thermodynamic potential).
The calculations are done in reduced variables:
$n_e/n_s=\lambda=3$, $\gamma=6$, $a_1=\ldots=a_4=a=0.015$, $\tilde
J=J/U=0.2$, $\tilde \mu= \mu/U =1$. Parameter $N=4$ in both
cases.}
\end{figure}

When the maximal number of bonds per molecule $N\geq4$, additional
solutions of equation (\ref{second}) arise. Mathematically, these
solutions arise due to kinks in the right hand side of
(\ref{second}), provided by high powers in the polynomial
$R(\Psi)$. As an example, the graphic solution of
eq.(\ref{second}) at some certain model parameters is presented at
fig.1c, together with corresponding $F(\Psi)$ (fig.1d) dependence.
As can be easily understood, the existence of such solutions may
result in the liquid - liquid phase transition. The model predicts
possible phase diagrams in the $\mu - T$ plane, which are
presented at fig.2. Phase diagrams of that type were suggested
earlier \cite{Malescio1,Malescio2}. Note, that the diagram at
fig.2b demonstrates an interesting possibility for the system
behavior: there is no first order transition in density parameter
$w$, and $\Psi$ occurs to be the relevant order parameter for the
liquid - gas critical point.

Thus, the mean - field approximation predicts two non - trivial
features. First, there exists metastable state with negative
$\Psi$. This state appears at rather low temperatures. Second,
high powers of $R(\Psi)$ which correspond to the case $N \geq 4$,
may provide liquid - liquid phase transition at some sets of
weights $a_n$. The transition is the first order phase transition,
which line terminates in the critical point at the $\mu - T$
plane. At this line, order parameter jumps from $\Psi_1$ to
$\Psi_2$. In the critical point, the jump vanishes. Since the
absolute value of $\Psi$ is proportional to the mean length of
quasi - polymers, then higher $\Psi$ corresponds to the higher
number of active bonds. The density difference is small, but
higher $\Psi$ corresponds to the higher density also.

To verify these results, in the next section we analyzed the exact
solution of the model on the Bethe lattice.

\begin{figure}[ht]\label{PD}
   \centering
 \includegraphics[width=0.52\textwidth,clip,viewport= 0 0 540 382]{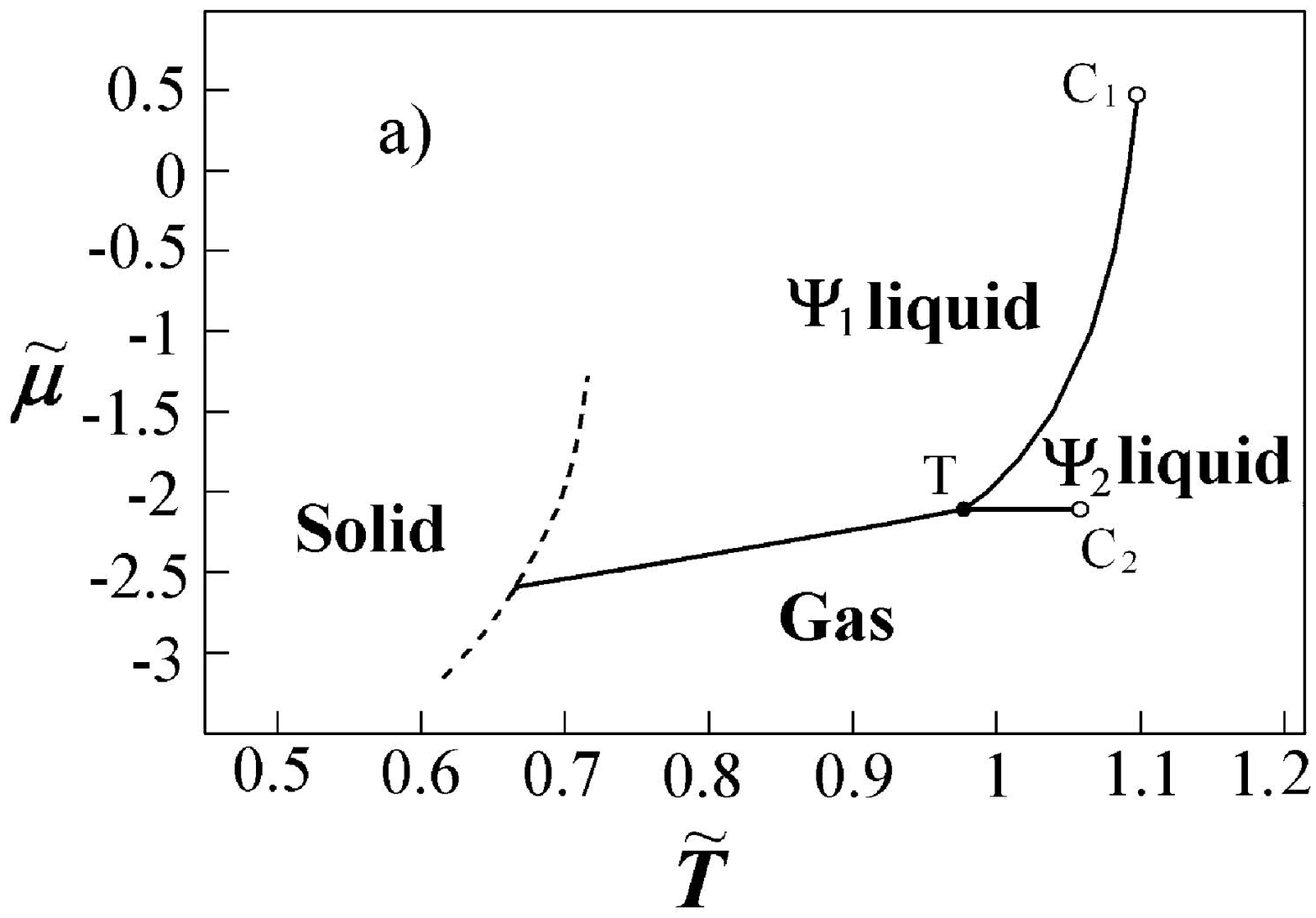}
  \hfil
      \includegraphics[width=0.46\textwidth,clip,viewport= 5 0 525 410]{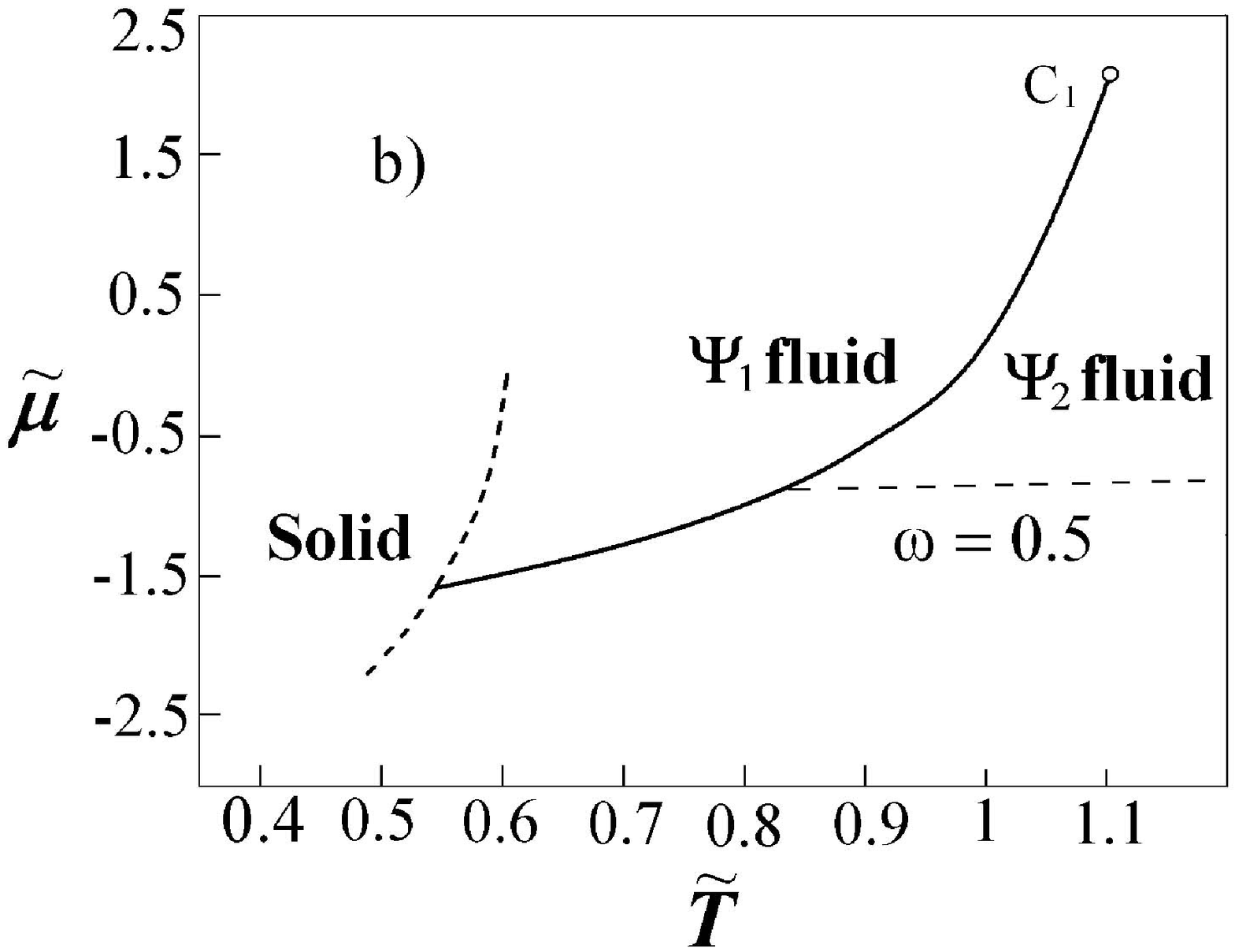}
     \caption{The $\tilde\mu - \tilde T$ phase diagrams at $\tilde J=0.7$
       (a) and $\tilde J=0.3$, (b). Other
      parameters are $n_e/n_s=\lambda=3,~\gamma=6,~a_1=\ldots=a_4=a=0.015,
       ~N=4$.
      Liquid - liquid and liquid - gas critical points are denoted
      as C$_1$ and C$_2$ respectively}
\end{figure}

\section{The model on the Bethe lattice.}

In this section, we put $\omega_\alpha=1$ for simplicity. All the
results obtained may be easily derived for arbitrary
concentration, but we are interested mostly in the high density
limit. Thus, we deal with the partition
\begin{equation}\label{Z3}
Z = \int {D\psi {\mathop{\rm e}\nolimits} ^{ - H_0 \left\{ {\psi
,s} \right\}} } \prod\limits_\alpha  {\left[ {1 + R_\alpha(\psi) }
\right]},
\end{equation}
on the Bethe lattice. Here,
$$
H_0 \left\{ {\psi } \right\} = \frac{1}{2}\sum_i K \psi _i^2
-\sum_i s_i \psi_i
$$
In $H_0$, we used the source term with $s_i=s$. Partition
(\ref{Z3}) may be treated as the generating function for mean
values
\begin{equation}\label{mean_psi}
\left\langle {\psi _i } \right\rangle  = \frac{1}{Z}\left.
{\frac{{\partial Z}}{{\partial s_i }}} \right|_{s = 0}
 ,\quad \left\langle
{\psi _i \psi _j } \right\rangle  = \frac{1}{Z}\left.
{\frac{{\partial ^2 Z}}{{\partial s_i \partial s_j }}} \right|_{s
= 0} ,\,\, \ldots.
\end{equation}
For each edge $i$ let us define function $Q(\psi_i)$, which is the
partition (\ref{Z3}) for sublattice, obtained from the initial
Bethe lattice by cut of the edge $i$. In the thermodynamic limit,
the partition for initial lattice depends no on the choice of the
edge and is given by
\begin{equation}\label{Q1}
Z = \int {d\psi _i e^{ - \frac{K}{2}\psi _i^2 +s_i\psi_i} } Q^2
(\psi _i ).
\end{equation}
Let us introduce the values
\begin{equation}\label{A}
A = \int {d\psi e^{ - \frac{K}{2}\psi ^2  + s\psi } } Q(\psi
)\psi,
\end{equation}
\begin{equation}\label{B}
B = \int {d\psi e^{ - \frac{K}{2}\psi ^2  + s\psi } } Q(\psi ).
\end{equation}
At $s=0$, these values are proportional to the probabilities of
the edge to be with or without bond, respectively. Obviously,
\begin{equation}\label{Q2}
Q(\psi ) = F_1 (A,B)\psi  + F_2 (A,B),
\end{equation}
where explicit form of functions $F_1(A,B)$ and $F_2(A,B)$ depends
on model details (the number of nearest neighbors and the set of
coefficients $a_n$). Integrating (\ref{Q2}), one gets system of
equations on values $A,B$:
\begin{eqnarray}\label{system_AB_1}
A = F_1 (A,B)J_2 (K,s) + F_2 (A,B)J_1 (K,s) \nonumber \\
{} \\
 B = F_1 (A,B)J_1 (K,s) + F_2 (A,B)J_0 (K,s)\nonumber.
\end{eqnarray}
Here,
$$
J_n (K,s) = \int\limits_{ - \infty }^{ + \infty } {d\psi e^{ -
\frac{K}{2}\psi ^2  + s\psi } } \psi ^n  = \sqrt {2\pi } K^{ -
\frac{3}{2}} \frac{{d^{n - 1} }}{{ds^{n - 1} }}\left(
{se^{\frac{{s^2 }}{{2K}}} } \right)
$$
From (\ref{mean_psi}), (\ref{Q1}), (\ref{Q2}),(\ref{system_AB_1})
one obtains
\begin{equation}\label{mean_psi_1}
\left\langle \psi  \right\rangle  = \frac{{2F_1 F_2 }}{{F_1^2  +
F_2^2 K}},\quad \left\langle {\psi ^2 } \right\rangle  =
\frac{{3F_1^2  + F_2^2 K}}{{F_1^2 K + F_2^2 K^2 }}.
\end{equation}
At $s=0$ system (\ref{system_AB_1}) takes the form
\begin{eqnarray}\label{system_AB_2}
A = F_1 (A,B)J_2 (K,0) = F_1 (A,B)\sqrt {2\pi } K^{ - \frac{3}{2}}
\nonumber \\
{} \\
 B = F_2 (A,B)J_0 (K,0) = F_2 (A,B)\sqrt {2\pi } K^{ - \frac{1}{2}}
 \nonumber,
\end{eqnarray}
and
\begin{equation}\label{mean_psi_2}
\left\langle \psi  \right\rangle  = \frac{{2AB }}{{KA^2  +
B^2}},\quad \left\langle {\psi ^2 } \right\rangle  = \frac{{3KA^2
+ B^2 }}{{K^2 A^2  + KB^2 }}.
\end{equation}
As can be seen from (\ref{system_AB_2}), (\ref{mean_psi_2}), any
mean value may be presented as a function of the parameter
$x=A/B$:
\begin{equation}\label{mean_psi_3}
\left\langle \psi  \right\rangle  = \frac{{2x}}{{Kx^2  + 1}},\quad
\left\langle {\psi ^2 } \right\rangle  = \frac{{3Kx^2  + 1}}{{K^2
x^2  + K}}.
\end{equation}
  The equation on  $x$ is
\begin{equation}\label{mean_x}
Kx=y(x),
\end{equation}
where $y(x)=F_1(x)/F_2(x)$.

Any thermodynamic function may be reconstructed using
$\left\langle {\psi ^2 } \right\rangle$. For example,
$$
\frac{{\partial \ln Z}}{{\partial T}} =  - \frac{U}{{2T^2
}}K\left\langle {\psi ^2 } \right\rangle.
$$
Using known relation for internal energy,
\begin{equation}\label{E}
E = T^2 \frac{{\partial \ln Z}}{{\partial T}}
\end{equation}
one gets
\begin{equation}\label{E(psi)}
E =  - \frac{U}{2}K\left\langle {\psi ^2 } \right\rangle.
\end{equation}
Inserting $F=-T \ln Z$ into (\ref{E}) and integrating, one gets
the free energy thermodynamic potential
\begin{equation}\label{F(T)}
F(T) - F(T_0 ) = \frac{T}{2}\int\limits_{T_0 }^T {d\tau
\frac{{K(\tau )}}{{\tau ^2 }}\left\langle {\psi ^2 (\tau )}
\right\rangle },
\end{equation}
where $\left\langle {\psi ^2 (\tau )} \right\rangle$ is defined by
(\ref{mean_psi_3}) and (\ref{mean_x}).

Thus, in the thermodynamic limit, any thermodynamic characteristic
may be calculated using $y(x)$. Now, let us determine this
function for arbitrary model parameters. Those are the maximal
number of bonds per molecule $N$,  the number of nearest neighbors
$\gamma$, and the set of coefficients $a_n$. The only limitation
imposed is $\gamma \geq N$. Consider some site $\alpha$ and
suggest that the integration in (\ref{Z1}) is done over all $\psi$
except those which correspond to the site $\alpha$. If one of
these edges is cut (for definition, the edge  $\psi_1$), then
initial lattice occurs to be divided in two sublattices (fig.3).

\begin{figure}[ht]\label{PD2}
   \centering
 \includegraphics[width=0.4\textwidth,clip,viewport= 10 5 355 370]{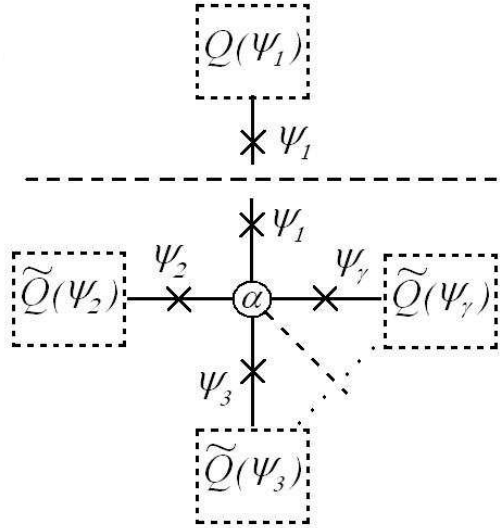}
       \caption{The cut of the initial Bethe lattice across the edge $\psi_1$
       leads to the equation (\ref{Q(psi_1)})}
\end{figure}

Then, $Q(\psi_1)$ may be written as
\begin{equation}\label{Q(psi_1)}
Q(\psi _1 ) = \int {\left( {\prod\limits_{i = 2}^\gamma  {d\psi _i
} e^{ - \frac{K}{2}\psi _i^2 } Q(\psi _i )} \right)} \left[ {1 +
a_1 (\psi _1  +  \ldots  + \psi _\gamma  ) + a_2 (\psi _1 \psi _2
+  \ldots  + \psi _{\gamma  - 1} \psi _\gamma  ) + \ldots }
\right].
\end{equation}
In the thermodynamic limit, $Q(\psi_i)$ depends no on $i$.
Integration together with (\ref{A}), (\ref{B}) and (\ref{Q2})
gives
\begin{eqnarray}
F_1 (A,B) = a_1 B^{\gamma  - 1}  + a_2 (\gamma  - 1)AB^{\gamma  -
2}  +  \ldots  + a_N C_{N - 1}^{\gamma  - 1} A^{N - 1}
B^{\gamma  - N} \nonumber \\
F_2 (A,B) = B^{\gamma  - 1}  + a_1 (\gamma  - 1)AB^{\gamma  - 2} +
\ldots  + a_N C_N^{\gamma  - 1} A^N B^{\gamma  - N - 1},
\end{eqnarray}
where $C^{n}_{m}=n!/m!(n-m)!$ - combinatoric coefficients. Since
$x=A/B$, $y(x)=F_1(x)/F_2(x)$, then
\begin{equation}\label{y(x)}
y(x) = \frac{{\sum\limits_{m = 0}^{N - 1} {a_{m + 1} C_m^{\gamma -
1} x^m } }}{{\sum\limits_{m = 0}^N {a_m C_m^{\gamma  - 1} x^m }
}}.
\end{equation}
Thus, equation (\ref{mean_x}) is algebraic with power  $N-1$.
Qualitatively, its structure coincides with mean - field equation
(\ref{second}). It should be mentioned, however, that
(\ref{mean_x}) may be treated as a recurrence relation
$Kx_{n}=y(x_{n-1})$, which connects $x$ values for $n$ and $n-1$
levels of the Cayley tree, taken in the thermodynamic limit. Since
any initial $x=A/B$ is positive, then all stationary points of
this recurrence are positive too. Thus, on the Bethe lattice, the
low - temperature metastable state does not exist.

\section{Conclusion}

Here, we point out some general results and features which make
the model rather attractive.
\begin{enumerate}
\item In mean field approximation, the metastable low - temperature
state is predicted. This state is universal, i.e. appears for any
set of model parameters. It becomes stable  only at  zero
temperature, but may be stabilized by an appropriate external
field. For example, phase separation in two phases with positive
and negative $\Psi$ may be achieved by impurity addition.
Recently, similar separation was observed experimentally
\cite{Hono}. For the Bethe lattice, the metastable state does not
exist. Since for the Bethe lattice closed paths are prohibited,
then it is natural to suggest that the metastable state may be
characterized by a set of self - closed quasi - polymers.
\item Liquid - liquid phase transition between phases with
different density of bonds is predicted both in mean - field
approximation and on the Bethe lattice. Possible phase diagrams
correspond to those predicted by computer simulations
\cite{Malescio1,Malescio2}. The phase transition is the first
order phase transition which takes place at some special sets of
model parameters. Also, the set of model parameters may provide
not phase transition but sharp crossover from one mean density of
bonds to another. In that case, the order parameter temperature
dependence demonstrates a kink. Thus, all $\Psi$ - dependent
properties should demonstrate a kink too. Such a kinks have often
been observed in molecular liquids \cite{lq1}-\cite{lq8}.
\item The model considered allows one to describe an arbitrary set of
molecules and their bonding. For example, different conformations
of bonds may be easily accounted by an appropriate choice of
$R(\psi)$.
\item Actually, the model considered is the vertex model. Such
models have often been used in statistical physics. For example,
Ising model may be formulated as the vertex model \cite{Baxter}.
Thus, the methods to investigate the model behavior are developed
rather well.
\end{enumerate}
 The work is supported by RFBR (grants N 04-02-96095,
 06-02-17269).

\end{document}